\documentclass[useAMS,usenatbib]{mn2e}

\usepackage[dvips]{graphicx}
\usepackage[]{txfonts}
\def\simless{\mathbin{\lower 3pt\hbox
{$\rlap{\raise 5pt\hbox{$\char'074$}}\mathchar"7218$}}}   
\def\simmore{\mathbin{\lower 3pt\hbox
{$\rlap{\raise 5pt\hbox{$\char'076$}}\mathchar"7218$}}}   
\newcommand{\be}{\begin{equation}}
\newcommand{\ee}{\end{equation}}
\title{Superflares from magnetars revealing the GRB central engine}

\author[Dimitrios Giannios]
{Dimitrios Giannios\thanks{E-mail: giannios@astro.princeton.edu}\\
Department of Astrophysical Sciences, Peyton Hall, Princeton
  University, Princeton, NJ 08544, USA\\}

\begin{document}
\date{Received / Accepted}
\pagerange{\pageref{firstpage}--\pageref{lastpage}} \pubyear{2009}

\maketitle

\label{firstpage}

\begin{abstract}

Long-duration gamma-ray bursts (GRBs) may be powered by the
rotational energy of a millisecond magnetar. I argue that the GRB-driving
magnetars lie at the high end of the distribution of
magnetic field strengths of magnetars.  The field of GRB magnetars decays
on timescale of hundreds of years and can power SGR-like flares up to $\sim 100$ times more
powerful than the 2004 event of SGR 1806--20. A few of these flares per year
may have been observed by {\it BATSE} and classified as short-duration GRBs. 
Association of one of these superflares with a nearby $d_L\simless 250$ Mpc galaxy 
and the discovery of a, coincident in space, 100-year-old  GRB afterglow 
(observed in the radio) will be the characteristic signature of the magnetar model for GRBs.

\end{abstract} 
  

\section{Introduction} 
\label{intro}

While long-duration GRBs have been shown to be associated with
core-collapse supernovae (Galama et al. 1998; Hjorth et al. 2003; Stanek et
al. 2003), the details on how the central engine of GRBs operates remain elusive.
The core of the star may collapse into a few solar mass black hole accretion
into which powers the GRB flow (Woosley 1993). Alternatively, a millisecond 
period protomagnetar may form at the stellar core. In this model, magnetic fields 
extract the rotational energy of the magnetar launching the
GRB flow (Usov 1992; Thompson 1994). 

Magnetars born with dipole surface fields $B_{\rm s}\sim 10^{14}-10^{15}$ G are common
making up around $\sim 10$\% of the neutron star population with a Galactic birth rate
of $\sim 10^{-3}$ yr$^{-1}$ (Kouveliotou et al. 1998). This rate is 2-3 orders of
magnitude higher than that of long-duration GRBs (corrected for beaming;
Guetta, Piran and Waxman 2005). If GRBs are connected to magnetar birth, a very 
small fraction of magnetars power GRBs indicating that special conditions 
need to be satisfied. 

From the theoretical perspective, fast rotation and very strong fields 
(even in comparison to those inferred for the Galactic magnetars)
appear to be needed for a successful magnetar model for GRBs (Thompson, Chang and
Quataert 2004;  Uzdensky and MacFadyen 2007; Metzger, Thompson and Quataert
2007; Bucciantini et al. 2009; see also Klu{\'z}niak and Ruderman 1998; 
Spruit 1999). Metzger et al. (2007) explored a variety of 
models for the proto-neutron star wind finding that $B\simmore 3\times 10^{15}$ G
and $P\simeq 1$ ms are required for a powerful wind of low baryon loading to
be launched within tens of seconds as needed to explain GRBs.

Galactic magnetars can release a substantial fraction of their magnetic energy during
powerful flares. The supergiant flare of the soft gamma-ray repeater (SGR)
1806--20 resulted in the release of $E_{\rm f}\sim 10^{46}$ ergs on a time scale
of $\sim 0.2$s (Palmer et al. 2005; Hurley et al. 2005; Frederiks et al. 2007; 
for a review see Mereghetti 2008). This energy corresponds to a
fraction $\sim 0.1 R_6^{-3}B_{15}^{-2}$  of the total magnetic energy contained in the neutron
star, where $R=10^6R_6$ cm is the radius of the star and $B=10^{15}B_{15}$ G is
the interior field strength. Such a flare could be detected up to a distance of 
tens of Mpc by {\it BATSE} (Palmer et al. 2005, Hurley et al. 2005).

Here, I argue that SGR flares from GRB-driving magnetars
are a factor of $\sim 100$ brighter than the December 2004 flare of SGR 1806--20
because of their tenfold stronger fields.    
Such flares should take place $\sim$hundreds of years after the formation
of the magnetar (and the associated GRB) and can be observed out to distance
of $\sim 250$ Mpc. I estimate that a few of these flares should be observed per
year. Radio observations at the location of the flare may be able to detect
and resolve the GRB afterglow proving the magnetar-GRB association.

\section{Superflares from GRB-driving magnetars}

The December 27, 2004 flare from SGR 1806--20 was extremely intense on Earth.
Saturation of the instruments makes the peak flux and spectrum of 
the flare hard to determine. Estimates for the peak luminosity of the 
flare range at $L_{\rm f}\sim 0.7\div 1.7\times 10^{47}$ ergs/s  with
the energy contained in the spike being $E_{\rm f}\sim 0.5\div 1.7\times
10^{46}$ ergs (assuming isotropic explosion and the revised distance 
of $d\simeq$9 kpc found in Bibby et al. 2008). The spectrum during the
flare is described by a power-law followed by an exponential cutoff slightly below $\sim 1$ MeV
(Palmer et al. 2005; Frederiks et al. 2007).  

SGR 1806--20 has estimated surface magnetic field of $B_{\rm s}\sim 8 \times 10^{14}$ G
within the typical range of the other known 
magnetars (Kouveliotou et al. 1998). For an interior magnetic field of 
$B\sim 10^{15}$ G, the total magnetic energy contained in the star 
(decay of which is believed to power the SGR activity) is 
\be
E_B=R^3B^2/6=1.6\times 10^{47}R_6^3B_{15}^2\quad \rm{erg}.
\ee 
Since the observed energy of the flare is not orders of magnitude less that
$E_B$, a substantial fraction (say 10\%) of the magnetic energy of the
magnetar can be released in a single event. 

The SGR and Anomalous X-ray Pulsar (AXP) activity in Galactic 
magnetars lasts for $\sim 10^4$ years
and is believed to be connected to the time it takes for the magnetic field
of the neutron star to decay. For magnetic fields in the magnetar
range, the field decay is mostly connected to ambipolar 
diffusion leading to (Thompson and Duncan 1996; Heyl and Kulkarni 1998)
\be
t_{\rm dec}\sim 10^4/B_{15}^2\quad \rm{yr},
\ee
that can be complicated by the cooling history of the neutron star
(not appearing explicitely in the last expression).

Successful models of magnetars as central engines for GRBs consider
core collapse to millisecond protomagnetars of $B\simmore 3\times 10^{15}$ G
(Metzger et al. 2007) with interior fields reaching strengths as high
as $B\sim 10^{17}$ G for substantial differential rotation of the 
protoneutron star (Klu{\'z}niak and Ruderman 1998; Spruit 1999)
\footnote{The fast rotation and very strong fields are likely connected since the 
$P\simeq 1$ ms rotation can lead to powerful magnetic fields 
through an efficient $\alpha-\Omega$ dynamo (Duncan and Thompson 1992; Thompson and
Duncan 1993).}. 
Under these conditions $\sim 10$s after core collapse the proto magnetar cools enough
for its neutrino-driven wind to weaken and a high-$\sigma$ GRB jet to be 
launched (Usov 1992; Thompson et al. 2004, Metzger et al. 2007).    
Assuming that the protoneutron star does not collapse into a black hole
shortly after the GRB, we are left with a magnetar of interior field of $B\sim 10^{16}$ G.
This field is expected diffuse from the magnetar on timescale of $\sim 100\div 1000$
yr (Heyl and Kulkarni 1998; ; Dall'Osso, Shore and Stella 2009; that is much
shorter of the typical Galactic magnetar lifetime) possibly powering SGR-type flares. 
Because of the stronger magnetic field at birth with respect to that of
a Galactic magnetar, the flares from GRB-driving magnetars
can be some 2 orders of magnitude more powerful than the August 
2004 flare of SGR 1806--20. The energy released in such ``superflares'' 
can be $E_{\rm sf}\sim 0.1 E_B\sim 10^{48}R_6^3B_{16}^2$ erg. Assuming duration of the superflares
of $\sim 0.1$s, their peak luminosity is $L_{\rm sf}\sim 10^{49}$
ergs/s.\footnote{If the duration of the SGR flares is connected to the 
time it takes for Alfv\'en waves to cross the neutron star (Thompson and Duncan
1996), the duration of the superflare can be shorter that $0.1$ s increasing its 
peak luminosity (for the same total energy released) and, therefore,
detectability. Such flares will fall within the short range of the duration
distribution of short GRBs.} Since not all the energy is used up in a single flaring 
a number of $f$ repetitions from the same source where $f\sim$ a few is also possible. 

\section{Predicted rates of superflares}

The distance out to which a flare from a GRB magnetar can be observed depends 
on both luminosity and spectrum. Assuming that 
the bulk of the energy of the flare is emitted at around $\sim 1$ MeV (as in
the 2004 flare of  SGR 1806--20), {\it BATSE} could detect such flare at a luminosity
distance of $d_L\simeq 250 L_{49}^{1/2}$ Mpc (e.g. Popov \& Stern 2006).
Guetta et al. (2005) estimate the  local ($z=0$) rate per unit volume 
of long-duration GRBs (corrected for
beaming) to be $R_{GRB}\sim 43 H_{71}^3 {\rm Gpc^{-3}yr^{-1}}$, where
$H_0=71H_{71}$ km/s/Mpc is the Hubble constant. If every long
GRB leaves a magnetar behind that powers $f$ superflares, the
rate of superflares within $d_L$ is\footnote{ This estimate assumes 
that the superflares are isotropic. If the flares are
beamed within a solid angle $\delta\Omega_{\rm sf}<4\pi$, only a fraction
$\delta\Omega_{\rm sf}/4\pi$ of the flares will be observed. On the other
hand, each flare will consume a factor $\delta\Omega_{\rm sf}/4\pi$
less magnetic energy from the magnetar potentially allowing for
a larger number of flares to occur. In this case $f$ stands for the number
of flares emitting within the observer's line of sight.}   
\be
{\dot R}_{\rm sf}=4\pi d_L^3 R_{\rm GRB}f/3=2.8fL_{49}^{3/2}\quad \rm{yr}^{-1}.
\ee 

This estimate shows that a few tens of the  
{\it BATSE} triggers over the 9 years of the mission
can be relatively nearby superflares.
These superflares would be classified as short GRBs. Out of the
$\sim 700$ short-duration GRBs observed by {\it BATSE} 
a fraction of $\sim 0.036fL_{49}^{3/2}$ may be flares from GRB-related
magnetars. 

The 2004 December 27 flare from SGR 1806--20 would have
been visible by {\it BATSE} out to tens of Mpc (Hurley et al. 2005;
Palmer et al. 2005). This triggered  several studies on the 
possibility that a fraction of short GRBs consists of
SGR flares {\it similar} to that of SGR 1806--20
(Lazzati,  Ghirlanda and Ghisellini 2005; Tanvir et al. 2005; Popov and Stern 2006; Nakar et
al. 2006; Chapman,  Priddey and  Tanvir 2009). Upper limits on this fraction of the order of $\sim 10$\%
were found by the lack of definite short GRB detections consistent
with the Virgo cluster (Palmer et al. 2005; Popov and Stern 2006), lack of possible 
hosts within 100 Mpc for six well localized GRBs (Nakar et al. 2006),
and the comparison with spectra of the brightest BATSE short GRBs (Lazzati et
al. 2005; hereafter L05). While these studies put constraints
on the fraction of Short GRBs originating within tens of Mpc, they cannot constrain
the possible contribution of superflares from GRB-related magnetars 
(discussed in this paper) to the {\it BATSE} sample.
Superflares are more rare, luminous and can be detected out to $\sim$hundreds of
Mpc. \footnote{L05 argue that no more that
$\sim$ 4\% of the BATSE short bursts are SGR flares, the reason being
that only few short GRBs are described by a thermal spectrum that
was inferred for the 2004 SGR 1806--20 flare (Boggs et al. 2007). However, 
the spectrum from the flare of SGR 1806--20 was hard to measure
because of saturation of the detectors and may actually be non-thermal (Palmer et al. 2005; 
Frederiks et al. 2007) making the L05 analysis less constraining. 
In any case our estimate on the superflares
``hidden'' in the short GRB sample is rather comparable to the 4\% upper limit
found by L05.}   
Interestingly, Tanvir et al. (2005) and Chapman et al. (2009)  find that some
$\sim 20$\% of the BATSE short GRBs with localization better than 10$^{\rm o}$
are correlated with galaxies out to $\sim 150$Mpc. A fraction of these bursts may
come from GRB-magnetar flares. 

\section{A magnetar flare revealing an old GRB afterglow}

Although an estimated a few long-duration GRBs per year take place within
$\sim 250$ Mpc (using the rates of, e.g., Guetta et al. 2005), 
we miss the prompt and early afterglow emissions from $\sim 99$\% of 
them because they are beamed away from us. About ten years after the GRB, 
however, the blast wave has decelerated to
sub-relativistic speeds and  the afterglow emission is 
bright at all directions. The magnetar flares can point to
the location where a long GRB {\it has already taken place some $\sim$100-1000
years ago} (the time it takes for $B\sim 10^{16}$ G fields to decay)\footnote{
The GRB can be accompanied with a supernova explosion. 
For typical parameters (e.g., Mazzali et al. 2003), the supernova ejecta remain Thomson thick 
for $\simless$ 3 years obscuring any early-time flaring from the magnetar but not 
the ($\simmore$ 100-year) late flares discussed here.}. 
Here, I show that the GRB afterglow emission should be still
detectable in the radio when the superflare takes place.

GRB afterglows can be followed in the radio wavelengths for years after the
burst. GRB 030329 is an intrinsically typical long GRB that took place particularly
nearby at $z=0.1685$ (or luminosity distance of $d_L=800$ Mpc for standard 
cosmology; Greiner et al. 2003). Its radio afterglow remains fairly bright (at the mJy level)
years after the burst and the blastwave is resolved (e.g. Berger et al. 2003; Taylor
et al. 2004; Resmi et al. 2005; Frail et al. 2005;  Pihlstr{\"o}m et al. 2007; 
van der Horst et al. 2008). Because of the slow decline in flux, the afterglow
is expected to be observable over the next decade in the $GHz$ range and
be {\it resolved} $\sim$7 years after the burst (Pihlstr{\"o}m et al. 2007).
With the Low Frequency Array ({\it LOFAR}) the afterglow of GRB 030329 can be detected for several
decades (van der Horst et al. 2008).

The afterglow emission of a GRB similar to that of 030329 
located at a distance $d_L\sim 250$ Mpc will be
$\sim 10$ times more bright and with the radio image a factor of
$\sim 2.6$ larger. Such an afterglow emission can be detected and 
resolved for hundred (or hundreds) of years after the burst.
Two-dimensional relativistic hydrodynamical simulations (Zhang and
MacFadyen 2009) indicate that the GRB blast reaches a distance of $\sim 3$ pc at $\sim 100$ years 
which corresponds to a source of angular size of 
$\sim 2.7$ mas (for a corresponding angular distance of $d_A\sim 224$ Mpc) 
and flux density of $\sim$0.1 mJy (at $\sim 1$GHz) allowing for the 
morphological study of the blastwave with high-sensitivity Very Long
Baseline Interferometry (VLBI) observations
similar to those reported in  Pihlstr{\"o}m et al. (2007).  
According to the same simulations,  the decelerating GRB blastwave is 
morphologically very different from a 
supernova remnant for the first $\sim 200$ years 
allowing for the distinction between the two types of explosions.       

For radio follow-ups to be possible, a good enough localization of the
superflare is needed. Such localization can be provided with the Burst
Alert Telescope ({\it BAT}) detector on {\it SWIFT}. The rate at which {\it SWIFT} detects GRBs is
$\sim$1/3 of that of {\it BATSE} mainly because of its smaller field
of view. I, thus, estimate that {\it BAT} detects
${\dot R}_{\rm sf}/3\sim 1fL_{49}^{3/2}$ superflares per year. {\it FERMI}
detection rate of flares is a factor of $\sim$2.5 higher but the 
Glast Burst Monitor ({\it GBM}) lacks
the localization needed for a radio follow-up. The pulsating tail that is
expected to follow the superflare may, in some cases, be powerful enough
to be observed with {\it XRT} hundreds of seconds after the event. Although
the pulsating tails that follow bright SGR flares of Galactic
magnetars for $\sim 200-400$s  have $L_{\rm tail}\sim 10^{42}$ ergs/s
(Mereghetti 2008), the strong magnetic field of the GRB magnetar can 
confine $\sim 100$ times more energy in the magnetosphere of the neutron star 
resulting in far brighter X-ray tails. It is furthermore possible 
that the superflare has a strong enough ``afterglow'' of its own that allows
for X-ray (or longer wavelength) detection and  accurate localization
shortly after the burst (Eichler 2002).

\section{Summary}

If GRB-magnetars exist, their magnetic field should decay
on a time-scale of a few hundred years possibly producing
SGR-like flares with peak luminosities of $\sim 10^{49}$ ergs/s.
A few of these flares per year should have been detected by 
BATSE out to $d_L\sim 250$ Mpc classified as short-duration GRBs.
Such superflares can be detected with {\it SWIFT} at 
a rate of about one per year. The host galaxy of the flare
should be typical of those of long-duration GRBs.   
High sensitivity radio observations at the location of the flare can resolve a 
$\sim 100$-year-old blastwave result of the interaction
of the GRB jets with the circumburst medium. This detection can
prove that GRBs are connected to the birth of magnetars.

\section*{Acknowledgments}
I thank Brian Metzger and Dmitri Uzdensky for stimulating discussions 
during the preparation of the manuscript.
I acknowledge support from the Lyman Spitzer, Jr. Fellowship awarded by the
Department of Astrophysical Sciences at Princeton University.

\end{document}